\documentclass[conference]{IEEEtran}

\usepackage{cite}
\usepackage{color}
\usepackage{multirow}
\usepackage{booktabs,multirow,array}
\usepackage[table]{xcolor} 
\usepackage{ltxcmds}
\usepackage{footnote}
\usepackage{amsthm}

\usepackage[table]{xcolor}
\usepackage{subfigure}
\usepackage{lipsum}
\usepackage{amssymb} 
\usepackage{pdflscape}
\usepackage{afterpage}

\usepackage{balance}

\usepackage{pifont}
\ifCLASSINFOpdf
\usepackage[pdftex]{graphicx}
\else
  \usepackage[dvips]{graphicx}
\fi

\begin{document}
\title{VeriBlock: A Blockchain-Based Verifiable Trust Management Architecture with Provable Interactions}

\author{\IEEEauthorblockN{Shantanu Pal$^{\ast}$, Ambrose Hill$^{**}$, Tahiry Rabehaja$^{\mathsection}$, Michael Hitchens$^{\mathsection\mathsection}$}
\IEEEauthorblockA{$^{\ast} $School of Computer Science, Queensland University of Technology, Brisbane, QLD 4000, Australia\\ $^{**}$Meadow Labs, Brisbane, QLD 4000, Australia \\ $^{\mathsection}$Risk Frontiers, Sydney, NSW 2065, Australia \\ $^{\mathsection\mathsection}$School of Computing, Macquarie University, Sydney, NSW 2109, Australia
\\
{shantanu.pal@qut.edu.au, ambrose@meadowlabs.io, tahiry.rabehaja@riskfrontiers.com, michael.hitchens@mq.edu.au} {}}}




\maketitle

\begin{abstract}
There has been considerable advancement in the use of blockchain for trust management in large-scale dynamic systems. In such systems, blockchain is mainly used to store the trust score or trust-related information of interactions among the various entities. However, present trust management architectures using blockchain lack verifiable interactions among the entities on which the trust score is calculated. In this paper, we propose a blockchain-based trust management framework that allows independent trust providers to implement different trust metrics on a common set of trust evidence and provide individual trust value. We employ geo-location as proof of interaction. Some of the existing proposals rely upon geo-location data, but they do not support trust calculation by multiple trust providers. Instead, they can only support a centralised system. Our proposed architecture does not depend upon a single centralised third-party entity to ensure trusted interactions. Our architecture is supported by provable interactions that can easily be verified using blockchain. Therefore, it allows a high degree of confidence in trust management by ensuring the actual interactions between the entities. We provide a detailed design and development of the architecture using real-world use case examples. The proof of prototype was implemented on the Ethereum blockchain platform. Experimental results demonstrate that the employment of independent trust providers adequately provides a high degree of trust scores and that the proposed architecture can be used in a real-world environment.
\end{abstract}


\IEEEpeerreviewmaketitle

\section{Introduction}
\label{introduction}
Trust management is fundamental in any computing system. According to the definition of~\cite{adewuyi2019ctrust}, a trust management system ``provides methods and mechanisms to evaluate the trustworthiness of interacting peers, based on a trust model''. It therefore helps transform choices into decisions while assisting in overcoming the uncertainty and risks associated with uncertain behaviour  secure data access. In a large-scale dynamic system, trust management is challenging due to the nature of the system, e.g., issues with the availability of information, reliability of entities providing the information, and a trustworthy entity that does not necessarily offer trustworthy information~\cite{pourghebleh2019comprehensive}. This further highlights the issue of how trust is assessed based on the available Quality of Information. Trust is generally calculated based on an entity's past behaviours or the interaction histories between two entities in a given context. The entity that is trusting is known as the \textit{trustee}, and the entity that is being trusted is known as the \textit{trustor}~\cite{pal2019towards}. In general, a question can arise whether a trustor can trust a trustee based on the provided information and, if so, in what aspect and to what extent. Significantly, trust does not automatically apply in both directions between trustee and trustor.

Trust is typically calculated from either or both of two different information sources: (i) \textit{direct} interactions, and (ii) \textit{indirect} interactions~\cite{guleng2019decentralized}. In the former, the trust is calculated based on the trustee's experience with an entity. In the latter, the trust is calculated based on a third party's experience with the entities. In both cases, the trust computation is significant for deciding on whether the trustee will interact with (or share information with) the trustor. For the indirect trust, the trustee is generally implicitly assuming the reliability of the third party reports of interactions. If such interactions can be proven to have transpired, then the basis on which the trust has been calculated becomes more substantial~\cite{pal2021internet}. 

Even if this problem with indirect trust can be solved, other issues remain. Most trust management frameworks that rely on a centralised architecture for trust management suffer from many disadvantages, e.g., central points of failure or system availability issues. They also show limited scalability. Moreover, they use a single, specific means of trust score calculation. This is, challenging for a large-scale dynamic system where the trust score calculation depends upon several factors, including the history of previous interactions, context, and the nature of interactions. This is even more complex when considering a system where multiple entities want to determine trust scores in numerous service providers for multiple resources~\cite{wei2022trust}. In a real-world, large-scale, and dynamic system, the number of interactions between the entities is potentially high and the uses to which trust is put equally varied. Therefore, it is unlikely that a single mechanism for calculating trust scores will suit the need of all entities. A large-scale system with multiple services providing trust scores from the same set of trust evidence is challenging~\cite{pal2018fine}. In this paper, we address these issues of the reliability of third-party reports and the question of providing multiple options for trust score calculation and provide an architecture to support such an approach using real-world use case scenarios. 
Further, we use geo-location data as proof of interaction. In our approach, we ensure verifiable interactions in a way that multiple independent trust providers and users can check them efficiently. Some of the existing proposals rely upon geo-location data, but they can not calculate trust by multiple trust providers in a decentralised fashion. Instead, they can only do so via a centralised system.

\begin{figure}
    \centering
    \includegraphics[scale=1.4]{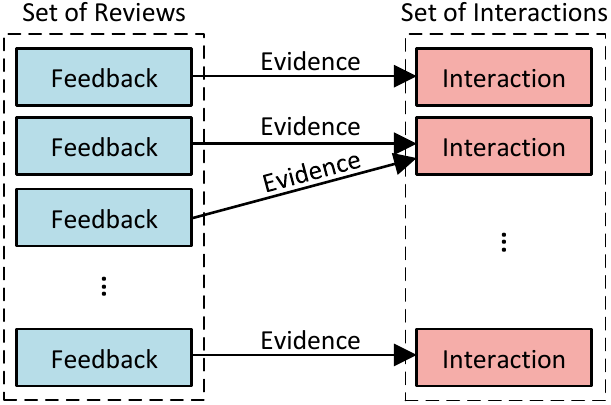}
    \caption{The relationship among reviews, feedbacks, interactions, and evidences of the proposed trust management architecture~\cite{PAL2021108506}.}
    \label{fig-1}
\end{figure}

In our previous work~\cite{PAL2021108506} we provided a mathematical model for the foundation of evidence-based trust with verifiable interactions. In this paper, we employ this model and propose an architecture that allows a group of \textit{trust providers} to provide trust scores independently, all using the same universal set of evidence. That is, we use a global set of evidence for the trust calculation as a universal basis. An \textit{evidence} is composed of an \textit{interaction} and an experience that constitute a \textit{review}. If the interaction indeed occurs (i.e., the interaction is proven), and the review pertains to that interaction, it becomes \textit{feedback}. A trust score is calculated from the identified feedbacks. In Fig.~\ref{fig-1}, we illustrate the relationship among reviews, feedbacks, interactions, and pieces of evidence. Significantly, our proposed trust management architecture holds two fundamental properties, (i) the availability of a universal set of evidence, and (ii) the existence of an evidence mapping that connects feedback to interactions. These properties enable the quantification of trust using evidence that can be traced back to interactions proven to have occurred, which is missing in the recent literature on trust management in large-scale dynamic systems.

We use blockchain to provide the basis of our trust management architecture~\cite{kumar2021leveraging}. Our intention in employing blockchain is to implement the two properties given above in a fine-grained manner. Blockchain is a distributed ledger for storing digital data in terms of transactions accessible across the entire network (i.e., to every entity within the network without a central authority). In addition, blockchain provides some salient features, e.g., transparent (all entities can see the copy of the ledger without disclosing their ownership), immutable (all validated records are cryptographically linked in a chronological sequence, which is impossible to change or reverse), time-stamped (each transaction is recorded within blockchain with a particular time-stamp), unanimous (all entities agree to validate a record), anonymous (the identity of interacting entities remain private), and programmable and logic-based processing (using smart contracts)~\cite{yue2021survey}. 

These features of blockchain ensure significant flexibility in our proposed architecture. For instance, the data stored on the blockchain becomes universal and transparent to all the entities within the network. The stored feedback information can not be altered or changed. This information can be verified based on a specific time-stamp. Furthermore, all entities in the blockchain can validate the same piece of information while remaining anonymous. Finally, smart contracts, held in the blockchain, help in the formation of the \textit{evidence-mapping}. In other words, blockchain helps in the verification of each feedback, i.e., that the interaction has occurred, and each feedback is linked to an interaction. Moreover, the trust providers can access the stored information on the blockchain. This effectively allows all trust providers (i.e., allowing multiple trust providers) to base their calculation of trust scores on the same evidence set, while carrying out the trust score calculations using different trust calculation algorithms. The major contributions of the paper can be summarised as follows: 

\begin{itemize}
    \item We propose an architecture for trust management in large-scale dynamic systems supported by verifiable interactions. Our architecture is supported by the provable interactions that can be verified using blockchain. This provides a high degree of confidence in trust management by ensuring the actual interactions between the entities.
    \item We provide a real-world use case scenario to demonstrate the feasibility of the proposed architecture. We also discuss the architectural design along with its components and communications and discuss the system functionality.
    \item We use geo-location data as proof of interaction for trust calculation. Unlike the existing proposals, in our approach, we acquire the verifiable interactions in a way that multiple independent trust providers and users can check them efficiently. 
    \item We examine the performance of the architecture using proof of prototype implementation on the Ethereum blockchain platform. To demonstrate the practicability, we use three different trust score mechanisms. The evaluation results show that the proposed architecture is efficiently feasible with different trust mechanisms. 
\end{itemize}

The rest of the paper is organised as follows. In Section~\ref{use-case}, we discuss a use case scenario. In Section~\ref{related-work}, we discuss the related works. Section~\ref{proposed-architecture} presents the proposed architecture along with the architectural components, system functionality, and communications between the various architectural components. In Section~\ref{implementation-evaluation}, we present the system design and implementation, and evaluate the achieved results. Finally, in Section~\ref{conclusion}, we conclude the paper. 

\section{A Use-Case Scenario}
\label{use-case}
In this section, we discuss a use case scenario to support our design: a smart vehicular transportation system. Suppose that cars (we postulate IoT-enabled smart cars) communicate with one another about real-time traffic information (e.g., notifying each other of heavy traffic or a roadside accident). This enables the recipient cars (i.e., the users in this case) to be better informed about the situation to make safer and more efficient decisions and produce more coordinated transport networks. However, the information shared among the cars needs to be trusted. A way to achieve this is through a collection of confirmation or rejection votes where each vote is proven to matter through the constraint of verifiable interaction.

\begin{figure}[t]
    \centering
    \includegraphics[scale=1.1]{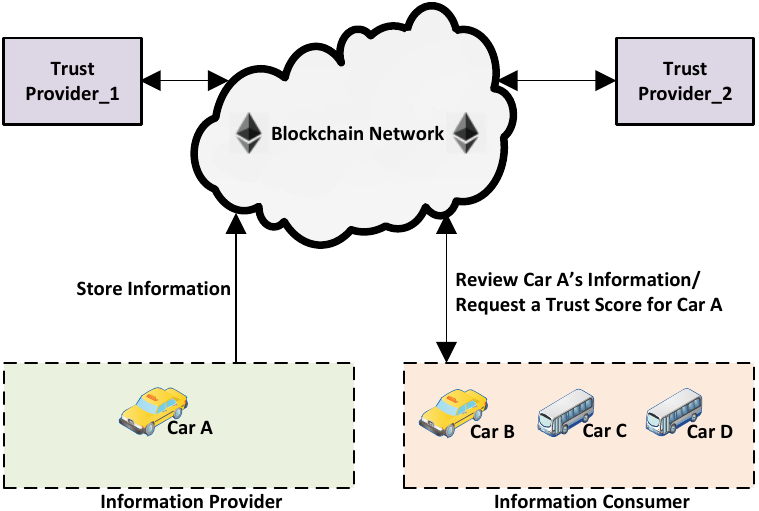}
    \caption{The use case scenario. Car A stores a piece of information on the blockchain, and the other cars (e.g., B, C and D) put reviews about this particular information supplied by Car A on the blockchain. There are many independent trust providers who provide trust scores for a particular information provider. Blockchain delivers a medium for communication among the information provider, information consumers, and trust providers.}
    \label{fig-use-case}
\end{figure}

In Fig.~\ref{fig-use-case} we illustrate the use case scenario. In the use case, a Car A can store a piece of traffic related information on the blockchain, and the other cars --- say B, C and D --- put reviews about this particular information on the blockchain. Therefore, both the information provided by Car A and the reviews performed by the other cars are stored on the blockchain. Car A can be seen as an \textit{information provider} and cars B, C, and D can be seen as the \textit{information consumers}. The blockchain provides a medium for communication between the information provider and consumers. In this example, we assume that the pertinent traffic information can be identified with the help of different sensors and automated systems such as cameras, LiDAR and potentially some clever use of mathematical and geospatial analysis. Some parts of such technology may still be emerging but prototypes are already used in self-driving cars  \cite{altaf2022survey}. The information is then transmitted and stored on the blockchain using a wide range of technologies, for example, WiFi or LoRa communication techniques \cite{haque2020lora}. In addition to the roadside information, a car's present position can also be captured by GPS (Global Positioning System), and it is stored on the blockchain linked to the submitted information or reviews. 

Let us assume that Car A has submitted some information about a roadside accident to the blockchain. Car B is passing by the same location and wants to submit a similar accident notice on the blockchain. However, it will notice (e.g., through a simple spatio-temporal buffer) that such information already exists, previously submitted by Car A. In this case, instead of submitting the same information once again, Car B submits a review of the information provided by Car A. Consequently, it will be a positive review for Car A. Note, in an adversarial condition (or being a selfish user) Car B can lie about the information in its review. However, in our approach, we can link the reviews to information usage as the cars pass through the same location around the same time. That is, we will have reviews with verifiable interactions. The trustworthiness of the information provided by Car A will then be based on the weighted sum of the number of cars providing the reviews, and a simple scoring can be to trust the majority. In a production system, we would assume a large number of reviews and part of the variance amongst trust providers would be the different means they adopt in dealing with false reviews. In contrast to Car B, Car C, which is currently parked at a different location, may be able to submit a review of whether the accident occurred or not, from a location that is far away from the actual place of the incident. In this case, our approach ensures that such a review be disregarded because the interaction cannot be verified. Thus the evidence mapping discussed in our previous work~\cite{PAL2021108506} can be built using a spatio-temporal constraint on the underlying information and reviews. This shows that the notion of verifiable interaction, implemented through the evidence mapping of Fig.~\ref{fig-1}, is more general than literal resource usage as per our previous work discussed in~\cite{PAL2021108506}. 

In this use case example, we assume that there are many independent \textit{trust providers} who provide trust scores for a particular information provider. These trust providers can filter in feedbacks, i.e., reviews with verifiable interactions, to be used for trust score calculations. Thus, if Car D wants to know whether the information from Car A is to be trusted, the trust providers can provide a trust score about Car A and the pertaining information. Car D can be sure that such a trust score was only computed from reviews left by cars which have travelled through the area near the accident location, and potentially previous scores of Car A. The trust providers retrieve evidence of interaction and calculate trust scores which, depending on the employed algorithm, can either be calculated in advance or on-demand. 

\section{Related Work}
\label{related-work}
There are several proposals that discuss blockchain-based trust management mechanisms. 
For instance, Chen et~al.~\cite{chen2020decentralized}, discuss a decentralised trust management architecture for ITS supported by blockchain. A consensus-based trust evaluation model is developed to store trust scores in a decentralised way. Blockchain is used as irreversible storage to hold the trust scores. 
Javaid et~al.~\cite{javaid2019drivman} present a trust management framework for vehicular ad-hoc networks (VANETs) using blockchain. Lahbib et~al.~\cite{lahbib2019blockchain} present a trust management architecture for large-scale dynamic systems (e.g., IoT) employing blockchain technology. Blockchain is used to share and store trust scores that enable a more reliable trust information integrity verification and enhanced data privacy. But how the evidence is linked to the reviews is not discussed in these papers. Further, unlike our proposal, they do not provide the foundation of evidence-based trust with verified interactions. 

Zhang et~al.~\cite{zhang2020blockchain} present a trust management system of IoV established on the blockchain. The system can detect vehicles that send malicious messages and penalise them by reducing their reputation values. Yang et~al.~\cite{yang2018blockchain} discuss a blockchain-based decentralised trust management system for vehicular networks. In this architecture, a vehicle can validate a received message from a vehicle at a close distance. Then based on the validation results, the vehicle renders a rating for an individual message (and therefore, for the source vehicle of the message). These ratings are stored inside a roadside unit that calculates the trust score and finally uploads on the blockchain. 

Dedeoglu et~al.~\cite{dedeoglu2019trust} discuss a blockchain-enabled trust management architecture for sensor nodes. The architecture provides end-to-end trust from data observation to blockchain validation. Further, to provide more granularity in trust management using observational data, the architecture considers the observer’s long-term reputation, level of confidence in its data, and verifies data from nearby observers. However, the actual data from observations is not stored on the blockchain. Instead, transactions that record the specific data and the reputation of nodes are stored on the blockchain. Unlike our approach, these proposals do not consider evidence-based trust with verifiable interactions between the information providers and information consumers, and do not address the significant issue of how the veracity of the information is assured.

In~\cite{singh2018blockchain},  Singh et~al. discuss a blockchain-based reputation framework for vehicular networks. The reputation helps build trustworthiness in a vehicle's communication in a specific geographical area with other vehicles. Blockchain is used to store each vehicle's reputation and public keys within the system. 
Hirtan et~al.~\cite{hirctan2020blockchain} present a similar approach of~\cite{singh2018blockchain}, for reputation and trust management for intelligent transportation systems using blockchain. The system decides the traffic conditions (e.g., traffic congestion or road maintenance works) within a specific area based on the geo-location data received by the vehicles passing through this area. Based on this data, alternate routes are determined for the other vehicles. 
Blockchain is used to store both reputation and valid geo-location traffic data. 
Yang et~al.~\cite{yang2019blockchain} propose a traffic event validation and trust management mechanism using blockchain. In this proposal, roadside units collect geo-location data from a specific geographical area and broadcast them to the other vehicles in this location. 
Although these proposals (i.e., \cite{singh2018blockchain},  \cite{hirctan2020blockchain}, and \cite{yang2019blockchain}) rely upon geo-location data, unlike our approach, they lack a way to independently calculate trust scores for multiple trust providers and multiple users.

In summary, we have noted that there has been significant consideration for employing blockchain for trust management. However, different from the existing works, our approach assures that an interaction indeed happened without relying upon a centralised component or readily observable behaviour. Moreover, our approach enhances the concept of multiple trust providers to access the evidence and provide individual trust values. Our approach is integrated with blockchain, leverages its several benefits, and allows the notion of trust-as-a-service.

\begin{figure*}
    \centering
    \includegraphics[scale=.7]{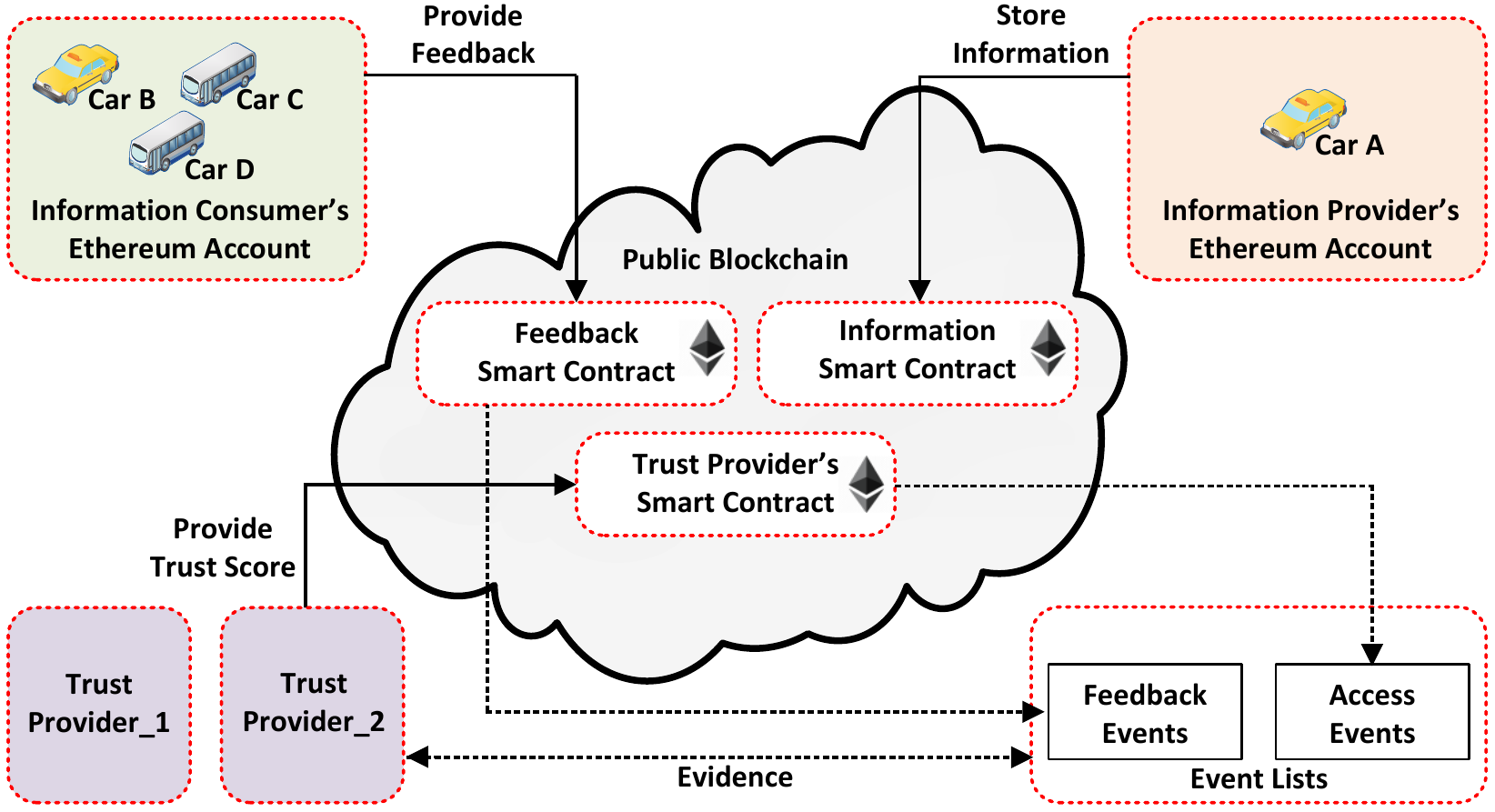}
    \caption{The proposed VeriBlock architecture. The \textit{`information smart contract'} controls access to certain information, the \textit{`feedback smart contract'} assures the reviews submitted by the information consumers, and the \textit{`trust provider's smart contract'} allows the trust providers to maintain trust scores. A \textit{`feedback event'} is associated with the user’s review (connects feedback to interactions), and feedback is linked to the \textit{`access event'} (ensures an interaction is verified).}
    \label{fig-architecture}
\end{figure*}

\section{VeriBlock: The Proposed Architecture}
\label{proposed-architecture}
In this section, we discuss the proposed VeriBlock architecture. We discuss (i) the different components of the architecture, (ii) system functionality, and (iii) show the communications between the architectural components.

\subsection{Components}
\label{components}
In Fig.~\ref{fig-architecture}, we present the architecture. We follow the architecture of our previous work in \cite{PAL2021108506}. Our proposed architecture is composed of the following components:

\begin{itemize}
    \item \textit{Information Provider:} An individual (or multiple) entity that provides information to other entities or organisations. An information provider maintains a set of information (or sets of information) collected, produced, or otherwise acquired from a real-world scenario and provides the information to others on demand.
    
    \item \textit{Information Consumer:} An individual (or multiple) entity that uses the information from other entities or organisations. An information consumer is capable of making decisions based on the available information.
    
    \item \textit{Trust Provider:} An entity that maintains the trust scores. They are responsible for executing the trust scoring functions and the outputs to the information consumers. In our approach, various trust providers can be implemented while accessing the same set of evidence. However, the implementation of such trust providers depends upon the choices of the individual designers of each provider.
    
    \item \textit{Blockchain Network:} A network of distributed digital ledgers that stores information across multiple computers. It is composed of a chain of blocks with a cryptographic hash value. Significantly, each block is linked to the other block's (i.e., previous block's) hash value~\cite{yue2021survey}. 
    Our design considers a public blockchain.
    
    \item \textit{Smart Contract:} A self-executing computer code stored on a blockchain enables the blockchain to automatically verify or intercede contract agreements between two parties under a set of conditions. Our design uses three smart contracts, (i) the \textit{information smart contract} that controls access to a piece of information, (ii) the \textit{feedback smart contract} that ensures the reviews submitted by the information consumers, and (iii) the \textit{trust provider’s smart contract} that allows the trust providers to maintain trust scores. We use these smart contracts to provide more granularity in secure access to restricted resources.
    
\end{itemize}

\subsection{System Functionality}
\label{system-functionality}

In this section, we discuss how the proposed architecture would function. Each car would be fitted with a portable IoT device capable of submitting blockchain transactions and signing with a private key. We assume this device is tamper-proof. The car would also have a set of sensors, including LiDAR, to autonomously detect information surrounding it (e.g., congested traffic conditions)~\cite{wang2017pedestrian}. Both information providers and information consumers will have this system available so that they can consistently communicate with the blockchain. Trust providers can interact with this network via their own hosted node or by accessing it through a public node. The nature of the blockchain (private or public blockchain) would also affect how each entity connects. In a public network, they would connect like any other entity, but blockchain node managers would need to be established in a private network. Then each car would need to be identified and given access to the network. For example, this could be done at the time of manufacturing, so car owners require no extra process.

As users of `VeriBlock' drive around, their cars will autonomously capture the data around them and report any anomalies, e.g., a car stopped on the side of the road, or multiple cars stopped in a highway lane. Information providers will provide the first reports of any incident. The car first detects an incident and then submits a transaction with the relevant information, time, location and classification. The blockchain will then emit an event that other cars can listen to. Other cars near that location can then submit feedback transactions to that event which will help improve the confidence of the system that an incident has actually occurred there. Trust providers can then calculate trust scores for each information provider that take into account the report or feedback location and confidence dependent on how many other vehicles support that evidence. We assume that the geo-location data is trustworthy. As the system continues to be used, these trust scores can be used to classify incidents faster and more accurately. With incidents being recorded securely and accurately, information consumers can start to make smart decisions around the route and take proper safety measures.
We only consider the interaction when the cars pass through the location and go in the same direction as the incident. So, for instance, if a car is on the other side of the road on a highway, it can still detect the accident, and it can register the incident. But as the car passes in the opposite direction, its feedback would not be counted. We also implicitly assume that the adversary can not spoof a location for a car. This is a separate area of research, and our work does not include solving this issue.

\subsection{Communication}
\begin{figure}
    \centering
    \includegraphics[scale=1]{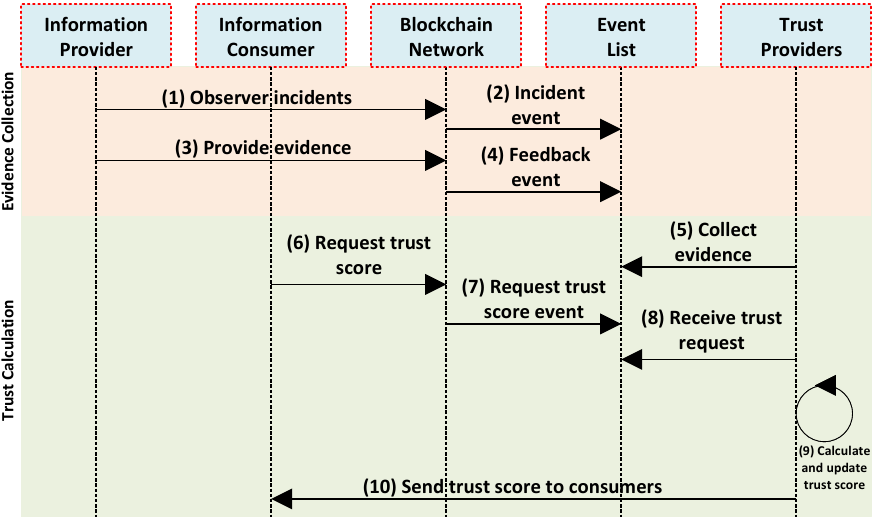}
    \caption{Communications among the different components of the proposed architecture. This is a more granular illustration of the elements and their interactions of Fig.~\ref{fig-architecture}.}
    \label{fig-communication}
\end{figure}

In Fig.~\ref{fig-communication} we illustrate the communication among the various components of our proposed `VeriBlock' architecture. 

\begin{itemize}[\label{}]
    \item Step 1: The cars that initially observe the incident (e.g., traffic congestion, etc.) will submit a transaction to the blockchain with the location and classification of the incident. We call such cars  information providers.
    \item Step 2: The blockchain network receives the initial transaction and submits an event (to the event list) notifying any listeners of the new incident.
    \item Step 3: Cars that are listening to the event list can then submit evidence or feedback for that particular incident. These pieces of evidence are time and location stamped which is used by the trust providers to filter out an accurate evidence.
    \item Step 4: The blockchain submits an evidence event for each piece of evidence submitted in Step 3. 
    \item Step 5: Trust providers can collect the evidence for each incident by listening to the event list and storing the particular evidence in a growing database. 
    \item Step 6: An information consumer can then request (and pay, if needed) for a trust score of a particular incident by submitting a request (and the payment) to the blockchain. 
    \item Step 7: The blockchain network will submit a request for a trust score event on behalf of the information consumer, while also holding any payment in escrow.
    \item Step 8: The trust provider receives the request by listening to the event list.
    \item Step 9: The trust provider calculates the trust score based on any trust algorithm they choose (cf. Section~\ref{results-discussion}). They calculate this score based on the data set collected in Step 5.
    \item Step 10: The trust provider then sends the updated trust score to the particular information consumer who requested this trust score. 
\end{itemize}

\section{System Design and Evaluation}
\label{implementation-evaluation}
In this section, we discuss (i) the implementation of the proposed  architecture, and (ii) evaluate the achieved results. 

\subsection{Implementation}
The implementation of the proposed `VeriBlock' architecture is split into two separate systems, (i) the blockchain network, and (ii) testing suite. The blockchain network is responsible for managing the data of the system, including storing, access control, and integrity. The testing suite is used to simulate car passing incidents and sending transactions to the blockchain. These systems in combination form the proof of concept for `VeriBlock'. 

The blockchain network consists of three smart contracts that are deployed on a private Ethereum network. The blockchain network consisted of a regular node and a miner. This was just to simulate roughly how the transactions would operate. These nodes were deployed on two different machines, (i) a 2016 Macbook Pro, that was set up as a regular blockchain node, and (ii) a 2021 Macbook Pro, that was setup as the miner of the private network. 

The testing suite was run on a Windows 10 Desktop and was built using Python 3.9 and the Web3 library to interact with the private Ethereum network. The transactions were created in batches of 1000 and sent to the network at random intervals. The transactions were set up in three different batches, a set where the feedback supports the initial observation, one where the feedback does not support the initial observation and a third set where each transaction is randomly split between positive and negative feedbacks. These three sets allow us to see the performance of the system under different stress loads and in different scenarios. 

The trust providers are also run on the Windows 10 Desktop. They were implemented in Python 3.9 as well. We implemented three different trust algorithms to show the versatility of `VeriBlock' at scale. We discuss this issue in more detail in the next section. 

\subsection{Results and Discussion}
\label{results-discussion}


The results show that the average time taken for a user (i.e., a car in this case) to submit a transaction and have it recorded on the blockchain is 12.3 seconds. The average time for a car to receive information from a request is around 6.3 seconds. The request of information is a simple read request, so it makes sense that this is the faster of the two operations. 

For the trust providers, the time of accessing the feedback data set is constant as again it is a read operation. The time of calculating the trust score for each observation depends upon the trust algorithm used. To demonstrate that our system can support trust providers implementing different trust algorithms, and that these will result in different trust scores based on the same evidence set, we use three different trust metric algorithms. While they are relatively simplistic algorithms, they do illustrate our proof-of-concept. The algorithms are: 

\begin{enumerate}
    \item Take all pieces of evidence for observation and treat each piece as equal. Use this to calculate the percentage of evidence that is positive (supports it). This percentage is then compared to a threshold to determine whether or not the observation happened. This threshold could be a fixed value or one that is nominated by the client on request. Note that this threat all evidence the same, there is no separation between direct and indirect trust. 
    \item Take all pieces of evidence for observation and treat them as equals. Use this to calculate the percentage of evidence that is positive. Then take all the evidence for that same observation but filter it on a metric, we used time and location to filter out more accurate evidence. Then average these two values and compare them to a threshold. 
    \item This is a variation on 2 above: instead of averaging the two values they are weighted, we used 70\% and 30\% for this example.
\end{enumerate}

\begin{figure}
    \centering
    \includegraphics[scale=.4]{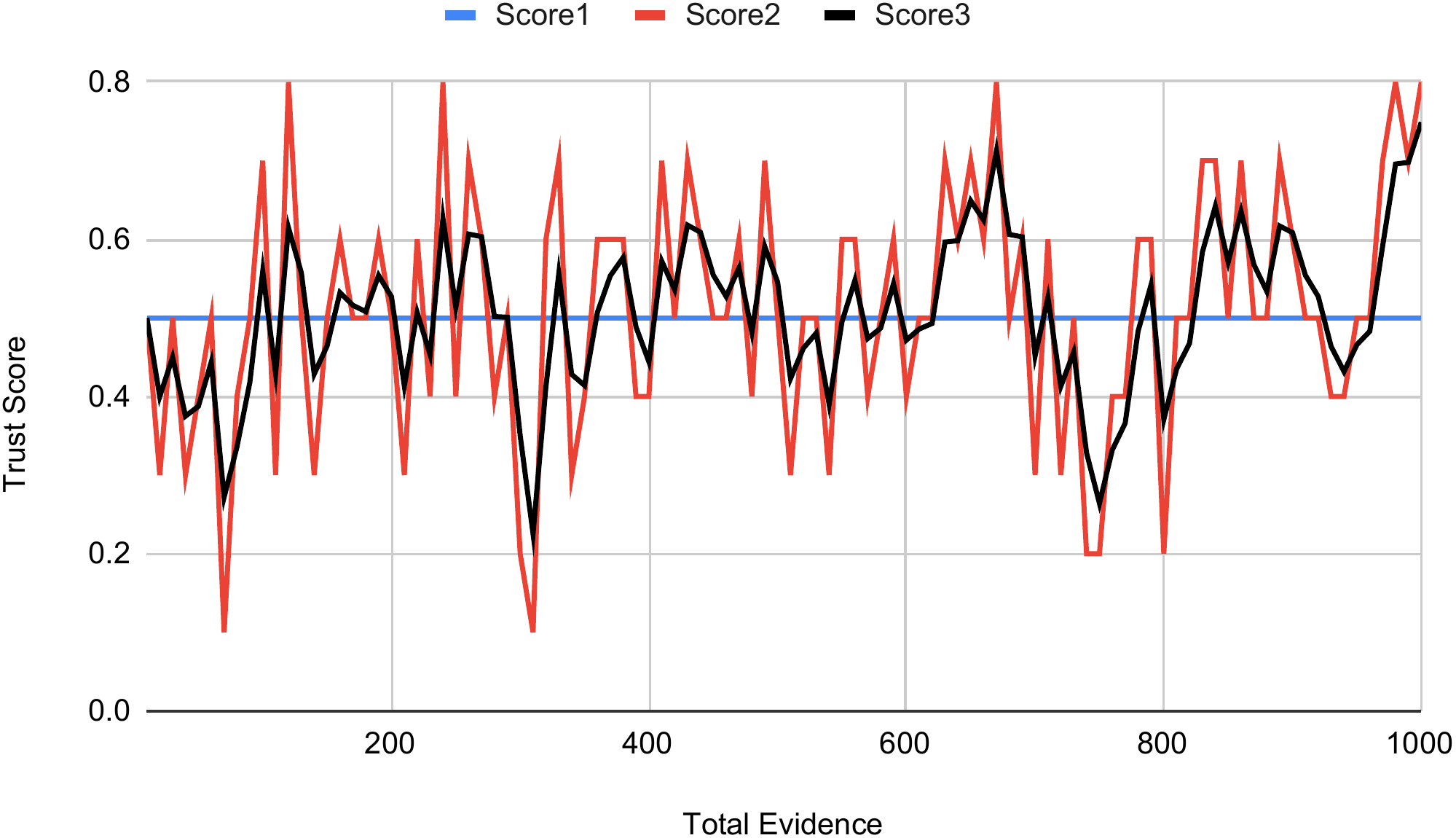}
    \caption{Results of different trust algorithms with 50\% good evidence.}
    \label{result1}
\end{figure}
\begin{figure}
    \centering
    \includegraphics[scale=.4]{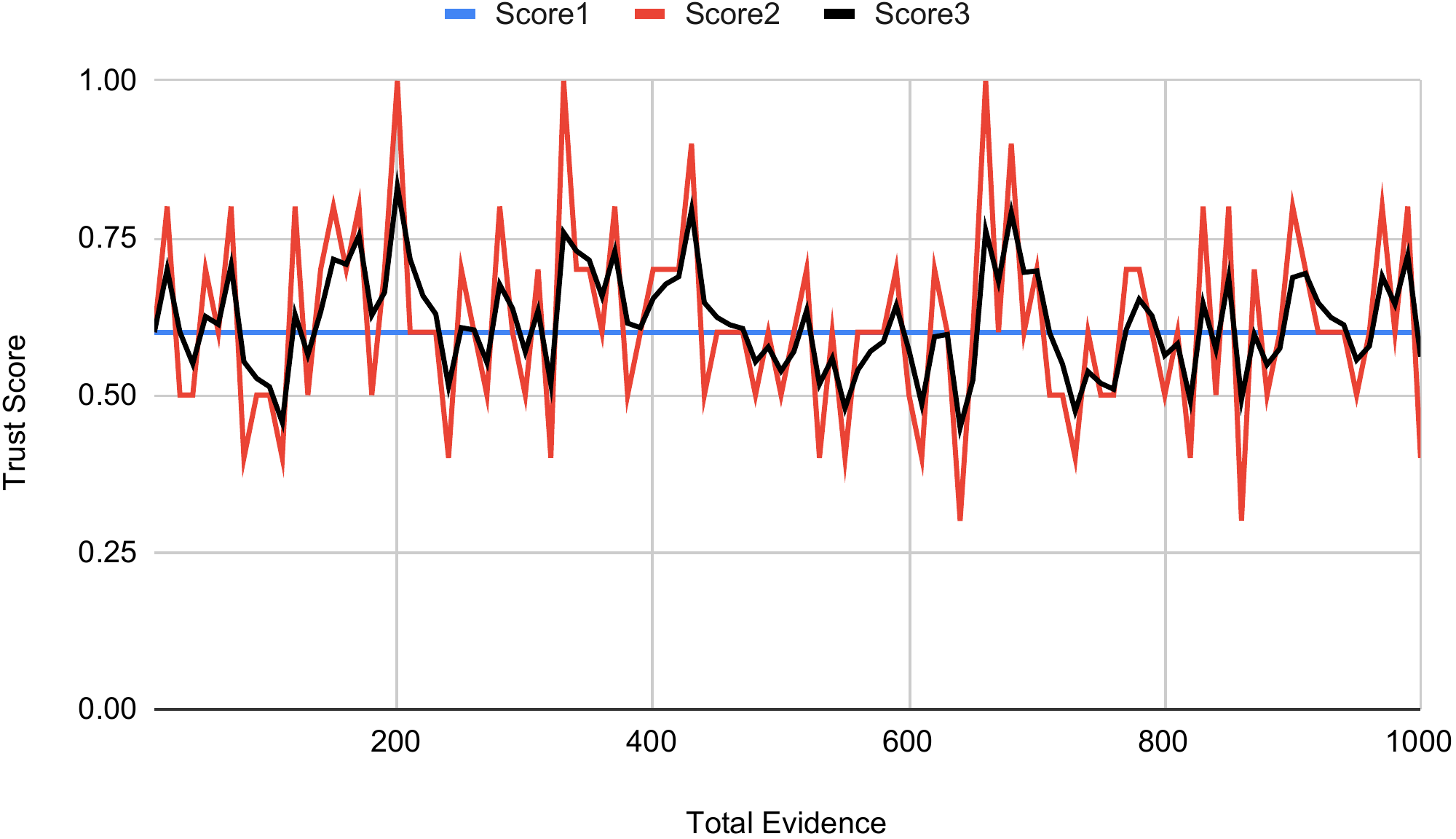}
    \caption{Results of different trust algorithms with 60\% good evidence.}
    \label{result2}
\end{figure}
\begin{figure}
    \centering
    \includegraphics[scale=.4]{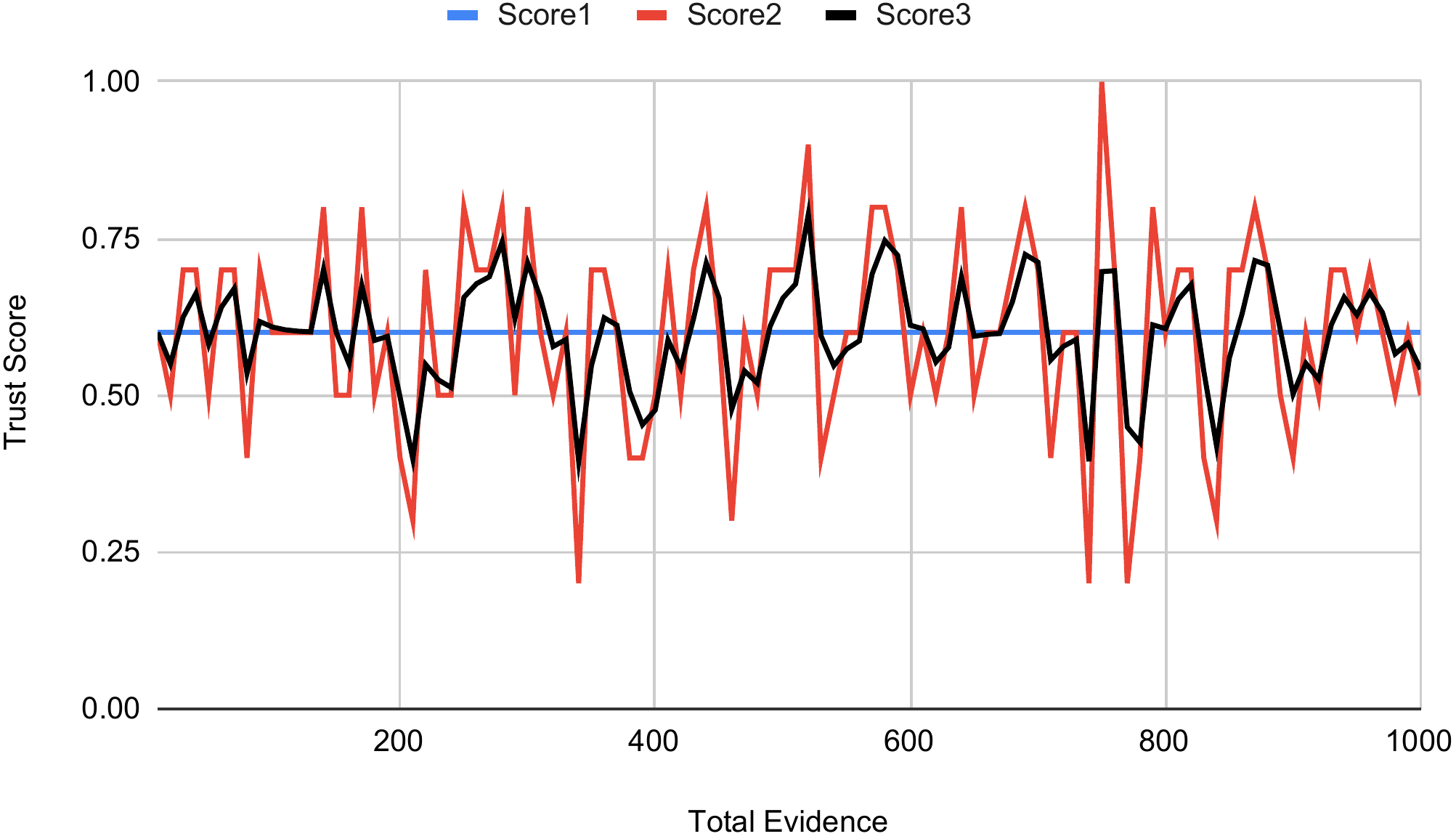}
    \caption{Results of different trust algorithms with 70\% good evidence.}
    \label{result3}
\end{figure}
\begin{figure}
    \centering
    \includegraphics[scale=.4]{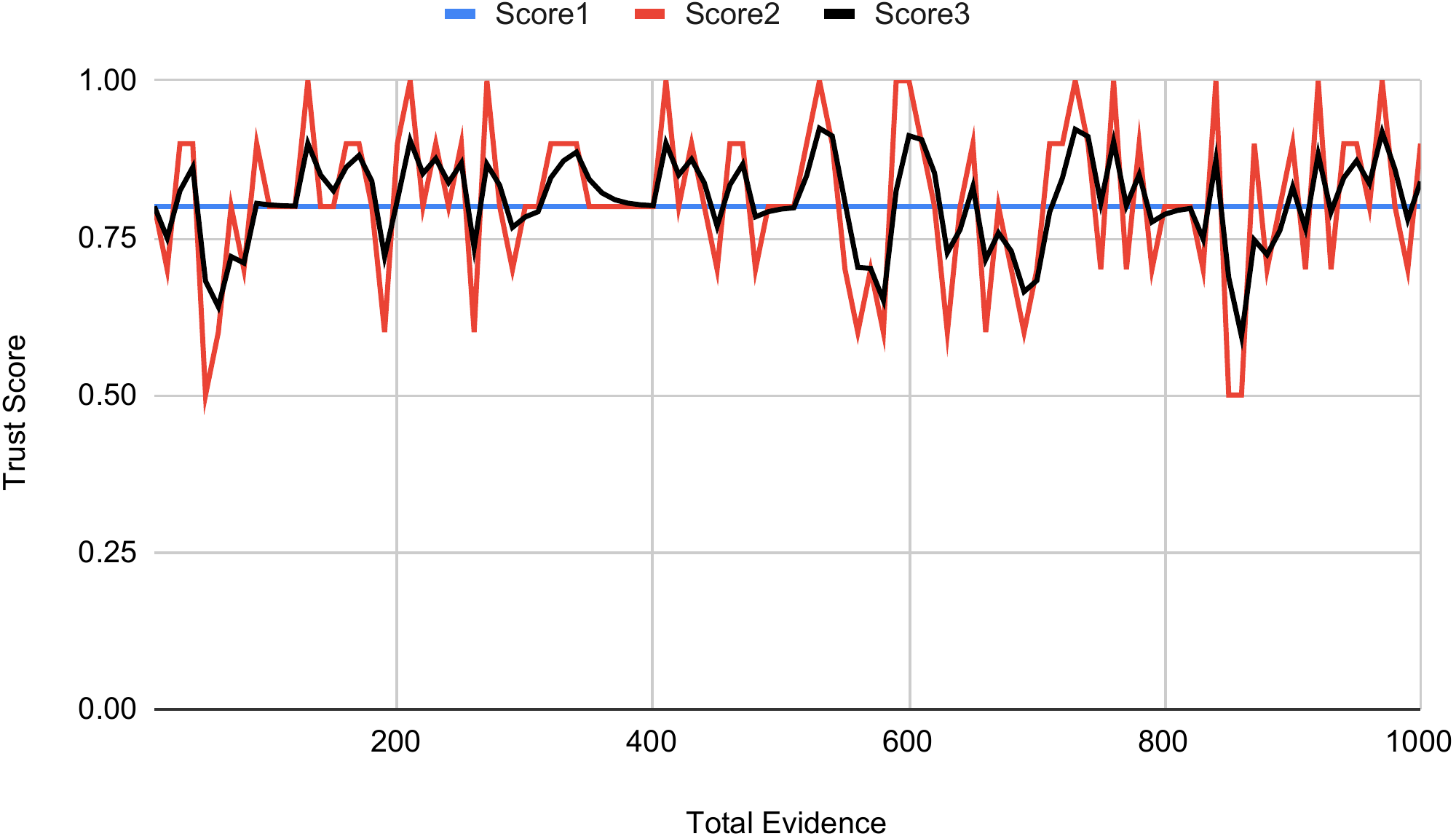}
    \caption{Results of different trust algorithms with 80\% good evidence.}
    \label{result4}
\end{figure}

For the trust score calculation: (i) we generate a set of ten pieces of evidence with a certain percentage being good. These percentages are chosen randomly, (ii) we calculate the three trust scores, (iii) we add another ten pieces of evidence with the same percentage being good, finally, (iv)  we repeat steps (ii) and (iii) above until 1000 total evidence has reached. We ran the test with 50\% (cf. Fig.~\ref{result1}), 60\% (cf. Fig.~\ref{result2}), 70\% (cf. Fig.~\ref{result3}), and 80\% (cf. Fig.~\ref{result4}) good evidence.
We note from the figures that our system supports trust providers using different trust metrics but drawing on the same body of evidence.

\section{Conclusion}
\label{conclusion}
In this paper, we have designed and developed a blockchain-based trust management architecture for large-scale dynamic systems. In our proposed architecture, pieces of evidence are backed up by provable interaction records. That is, the architecture can determine the verified interactions between the entities for a trust score calculation. We used multiple trust providers that can individually provide trust value on the same set of evidence. This improved the trust calculation procedure based on the actual set of evidence in highly dynamic and decentralised systems. We developed a proof of concept prototype implementation based on the Ethereum blockchain. Our experimental results demonstrated that the architecture is capable of managing multiple trust providers efficiently. We plan to explore a large-scale simulation with other use case scenarios in the future.

\ifCLASSOPTIONcaptionsoff
  \newpage
\fi


\bibliographystyle{IEEEtran}
\bibliography{bare_jrnl}

\end{document}